\documentclass[12pt]{report}

\usepackage{graphicx}
\usepackage{epsfig}

\usepackage{times}

\begin{document}

\setcounter{page}{109}
\setcounter{chapter}{6}

%\documentclass{book}
%%\documentclass{article}
%%\documentclass[11pt]{article}
%\usepackage{epsfig}
%\textwidth 11.5cm
%\textheight 18.5cm \topmargin -.5cm
%\textheight 26.5cm \topmargin -2.5cm
%\oddsidemargin 0.6cm
%%%%%%%% \usepackage{times}

%\begin{document}

%\chapter%
%[Poincar\'{e}-based 
%chaos 
%control of delayed measured systems: Limitations \& Improved Control]
%\mbox
\mbox{}\\[24mm]
\noindent {\Large\sf 6 
% ~~~~~~~~~~~~~~~~~~~~~~~~~~~~~~~~~~~~~~~~~~~~~~~~~~~~~~~~~~~~~~~~~~~~~~~~~~~~~~~~~~~~~~~~~~~~~~~~~~~~~~~~~~~~~~~~~~~~~~~~~~~ 
\mbox{}}
\\[2mm]
%\chapter
{
%\mbox{
%
\Large
\sf Poincar\'{e}-based
% chaos 
control of delayed measured systems:
%}
%\\
\\[2mm]
%\mbox{
\Large\sf Limitations and Improved Control%
%}
\footnote{To
appear in:
%Eckehard Sch\"oll and Heinz Georg Schuster
E.\  Sch\"oll \& H.\ G.\ Schuster 
(Eds.),
Handbook of Chaos Control, 
Wiley-VCH (2007).
}
} 
\mbox{}\\\\\\
\noindent {\large\sl Jens Christian Claussen}
%\\\\
%\chapterauthor[Jens C. Claussen]{Jens Christian Claussen}
%\chapter*{\LARGE \bf 1. Poincar\'{e}-based chaos control of  delayed  measured systems:\\ Limitations and Improved Control} \mbox{} \\ \\ 
%\small
%\footnotesize
%Institut f\"ur Theoretische Physik und Astrophysik,
%\\
%Christian-Albrechts-Universit\"at zu Kiel
%\\
%Leibnizstr.\ 15, D-24098 Kiel, Germany
%B\\
%\verb+claussen@theo-physik.uni-kiel.de+
\normalsize

\newcommand{\IGN}[1]{{\sl #1}}

\newcommand{\Nullop}{{\bf 0}}
\newcommand{\Einsop}{{\bf 1}}

\section{Introduction}
\index{Measurement delay}
\index{Delayed measurement}
\index{Delayed knowledge}
\index{Delayed feedback}
\index{Control loop latency}
\index{Poincare based control}
What is the effect of measurement delay on 
Ott, Grebogi, and Yorke (OGY)
 chaos control?
Which possibilities exists for improved control?
These questions are addressed within this chapter, and
the OGY control case is considered as well as a related
control scheme, difference control; both
% taken 
together form 
the two main Poincar\'e-based
chaos control schemes, where the control amplitude is 
computed once during the orbit after crossing the Poincar\'e section.

%In the case that
If the stabilization of unstable periodic orbits or fixed
points by the method given by 
Ott, Grebogi,
% and 
Yorke
% (OGY) 
\cite{ogy90}
and
% similarily
 H\"ubler 
\cite{hubler}
can only be based on
a measurement delayed by $\tau$ orbit lengths, 
resulting in a
% respective
control loop latency,
the performance of unmodified
OGY control is expected to decay. 
For experimental considerations, it is
desired to know the range of stability with minimal knowledge of the system.
In section 
%\ref{sec_unmodif},
\ref{sec_ogy_etc},
the area of stability is investigated
both for OGY control  and for difference control,
yielding a delay-dependent maximal Lyapunov number beyond which control fails.
%Hence, both unmodified Ott-Grebogi-Yorke control and difference control can be successfully applied only for a
%certain range of Lyapunov numbers depending on the delay time.
Sections
\ref{sec:rhythmic} to \ref{sec:mdc}
address the question how the control of delayed
measured chaotic systems can be improved, i.e., what extensions must be
considered if one wants to stabilize fixed points with a higher Lyapunov number.
% $\lambda_{\rm{}max}$. 
Fortunately, the limitation can be overcome most elegantly by
linear control methods that employ memory terms, as
linear predictive logging control 
(Sec.\ \ref{sec:lplc})
and memory difference control (Sec.\ \ref{sec:mdc}). 
\index{deadbeat control}
In both cases, one is equipped with 
an explicit deadbeat control scheme that
allows, within linear approximation, to perform control without principal
limitations in delay time, dimension, and Ljapunov numbers.

\subsection{The delay problem - time-discrete case}
For fixed point stabilization
in time-continuous control,
the issue of delay has been investigated
widely in control theory,
dating back at least to
the Smith predictor
\cite{smith57}.
\index{Smith predictor}
This approach mimics the, yet unknown, actual
system state by a linear prediction based on the
last measurement.
Its time-discrete counterparts discussed 
in this chapter
allow to place all eigenvalues of the associated
linear dynamics to zero, and always ensure stability.
The (time-continuous) Smith predictor
with its infinite-dimensional initial condition
had to be refined \cite{palmor80,hagglund96}, 
giving rise to the recently active fields of 
{\sl model predictive control}
% or {\sl receding horizon control}
\cite{mpc}.
For fixed point stabilization, an extension 
of permissible latency has been found
for a modified proportional-plus-derivative controller
\cite{sieber04}.

Delay is also a generic problem in the control of chaotic systems.
The effective delay time $\tau$ in any feedback loop is the sum of at least 
three delay times, the duration of measurement, the time needed to compute 
the appropriate control amplitude, and the response time of the
system to the applied control. The latter effect appears especially when
the applied control additionally has to propagate through the system.
These response time may extend to one or more cycle lengths
\cite{mausbach95}.
If one wants to stabilize the dynamics 
of a chaotic system onto an unstable periodic orbit,
one is in a special situation.
In principle, a proper engineering approach could be 
to use the concept of sliding mode control
\cite{slidingmode},
i.e.\ to use a co-moving coordinate system
and perform suitable control methods within it.
However, this requires quite accurate knowledge of
% the 
whole trajectory 
and 
%the local derivatives of the dynamical flow, to locate
% the direction of the
stable manifold,
with
% the
respective numerical or experimental costs.

Therefore direct approaches have been developed
by explicitely taking into account either
a Poincar\'e surface of section \cite{ogy90}
or the explicit periodic orbit length \cite{pyragas92}.
This field of {\sl controlling chaos}, 
or stabilization of chaotic systems, 
by small perturbations,
in system variables \cite{hubler}
or control parameters \cite{ogy90},
has developed to 
%emerged to
 a widely discussed topic with applications
in a broad area from technical to biological systems.
Especially in fast systems 
\cite{socolar94,blakely}
or for slow drift in parameters \cite{Cla98,mausbachpp99},
difference control methods have been successful,
namely the time-continuous Pyragas scheme
\cite{pyragas92},
ETDAS \cite{socolar94},
and time-discrete difference control
\cite{bielawski93a}.

\pagestyle{headings}

Like for the control method itself, 
the discussion of the measurement delay
problem in chaos control 
has to take into account the special issues of the situation:
In classical control applications
one always tries to keep the control
loop latency as short as possible.
In chaotic systems however, 
one wants to control a fixed point 
of the Poincar\'e iteration
and thus has to wait until the next
crossing of the Poincar\'e surface of section,
where the system again is in vicinity of that fixed point.

The stability theory and the 
delay influence for time-continuous chaos control schemes
has been studied extensively
\cite{just97,just99pla,franc99,just99pre,hovel}, 
and an improvement of control by periodic modulation 
has been proposed in \cite{just03pre}.
For measurement delays that extend to a full
period, however no extension of the time-continuous 
Pyragas scheme is available.

This chapter discusses 
the major
Poincar\'e-based
control schemes
OGY control \cite{ogy90}
and difference feedback \cite{bielawski93a}
in the presence of time delay,
and addresses the question what strategies
can be used to overcome the limitations
due to the delay
\cite{Cla04}.
We show how the measurement delay
problem can be solved systematically
for OGY control and difference control
by rhythmic control and memory methods
and give constructive direct and elegant formulas 
for the deadbeat control in the time-discrete
Poincar\'e iteration.
While the predictive control method LPLC presented below 
for OGY control
has a direct correspondence to the Smith predictor
and thus can be reviewed as its somehow straightforward
implementation within the unstable subspace of the
Poincar\'e iteration,
this prediction approach does not guarantee a 
stable controller for difference control.
However, within a class of feedback schemes
linear in system parameters and system variable,
there is always a unique scheme where all eigenvalues
are zero, i.~e.\  the MDC scheme presented below.
The method can be applied also for more than one positive Ljapunov
exponent, and shows, 
within validity of the linearization in vicinity of the orbit,
to be free of principal limitations in Ljapunov exponents or delay time.
For zero delay (but the inherent period one delay of difference control),
MDC has been demonstrated experimentally 
for a chaotic electronic circuit \cite{Cla98}
and a thermionic plasma discharge diode \cite{mausbachpp99},
with excellent agreement, both of stability areas
and transient Ljapunov exponents, to the 
theory presented here.
%
%\IGN
%{\sl
This chapter
 is organized as follows. 
After introducing the notation within a recall
of OGY control, 
we give a brief 
summary what limitations occur for unmodified
OGY control; 
details can be found in \cite{Cla04}.
From Section \ref{sec:memoryx}
we introduce different memory methods to improve control,
of which the LPLC approach appears to be superior as
it allows stabilization of arbitrary fixed points
for any given delay.
The stabilization of unknown fixed points is discussed in
Section 
%\ref{sec:diffkont} and
 \ref{sec:mdc}, 
where we present a memory method 
(MDC) that again allows stabilization
of arbitrary unstable fixed points.
For all systems with only one
instable Lyapunov number, the iterated dynamics can be transformed
on an eigensystem which reduces to the one-dimensional case,
and the generalization to the case of higher-dimensional
subspaces is straightforward
\cite{ClaSchu04}.
%}%IGN#
%
%
%
%
%
%

%\clearpage
\subsection{Experimental setups with delay}
\index{Poincare section}
Before discussing the time-discrete reduced dynamics 
in the Poincar\'e iteration, it should be clarified
how this relates to an experimental control situation.
On a first glance, the time-discrete viewpoint seems to
correspond only to a case where the delay 
(plus waiting time to the next Poincar\'e section)
 exactly matches the orbit length,
or a multiple of it. 
%%The generic experimental  situation however comes up with
Generically, in the experiment one
experiences
 a non-matching delay.  
Application of all control methods discussed here 
requires to introduce an additional delay,
usually by waiting for the next Poincar\'e crossing,
so that measurement and control are applied 
without phase shift at the same position of the 
orbit. 
In this case the next Poincar\'e crossing
position $x_{t+1}$ is a function of 
the values of $x$ and $r$ at a finite number of
previous Poincar\'e crossings only, 
i.~e. it does not depend on intermediate positions.
Therefore the (a priori infinite-dimensional)
delay system reduces to a finite-dimensional
iterated map.
If the delay (plus the time of the
waiting mechanism to the next Poincar\'e crossing)
does not match the orbit length,
the control schemes may perform less efficient.
Even for larger deviations from the orbit, the
time between the Poincar\'e crossings
will vary only marginally, thus a
control amplitude should be available in time.
In practical situations therefore the
delay should not exceed the 
orbit length minus the variance of the
orbit length that appears in the respective
system and control setup.

In a formal sense, 
the Poincar\'e approach 
ensures robustness with respect to 
uncertainties in the orbit length,
as it always ensures a synchronized reset
of both trajectories and control.
Between the Poincar\'e crossings the control
parameter is constant, the system is independent of
everything {\sl in advance of} the
last Poincar\'e crossing.
It is solely determined by the 
differential equation (or experimental dynamics).
Thus the next crossing position is
a well defined iterated function of the 
previous one.
This is quite in contrast to the situation of a
delay-differential equation (as in Pyragas control),
which has an infinite-dimensional initial condition
it `never gets rid of'.
One may proceed to stability analysis via
Floquet theory 
\cite{hale}
as investgated for
% various 
continuous \cite{just97}
and 
impulse length issues in
Poincar\'e-based 
\cite{claussenthesis,claussenfloquet,claussenfloquetenoc} 
control schemes.
\index{Impulse length}
Though a Poincar\'e crossing detection may 
be applied as well, the position will depend not only 
on the last crossing, but also on 
all values of the system variable within a time horizon 
defined by the maximum of the delay length
and the (maximal) time difference between two Poincar\'e crossings
(being non-stroboscopic).
Thus the Poincar\'e iteration would be a
function between two infinite-dynamical spaces. 
%Apart from further mathematical subtleties,
Contrary to
a delay differential equation with 
{\sl fixed} delay,
a major advantage of a Poincar\'e map
is to reduce the system dynamics
to a low-dimensional system;
therefore
\IGN{for all control schemes discussed here,}
the additional dimensionality is not
a continuous horizon of states, but merely
 a finite
set of values
 that were
measured at
 the 
previous Poincar\'e crossings.

\section{Ott -- Grebogi -- Yorke (OGY) control}
\index{OGY control}
% \section{Control of unstable periodic orbits and delayed measurement}
The method of Ott, Grebogi and Yorke
\cite{ogy90}
stabilizes unstable fixed points, 
or unstable periodic orbits
utilizing a Poincar\'e surface of section,
by feedback that is applied in vicinity of the fixed point $x^*$ 
of a discrete dynamics $x_{t+1}=f(x_t,r)$.
For a chaotic flow, or corresponding experiment,
the system dynamics
$\dot{\vec{x}} =\vec{F}(\vec{x},r)$
reduces to the discrete dynamics between subsequent
Poincar\'e sections at $t_0, t_1, \ldots t_n$.
This description is fundamentally different from a stroboscopic sampling
as long as the system is not on a periodic orbit, where the
sequence of differences $(t_i-t_{i-1})$ would show a periodic structure.

If there is only one positive Ljapunov exponent,
we can proceed considering the motion in unstable
direction only.
One can transform on the eigensystem of the Jacobi matrix
$\frac{\partial{} f}{\partial{} r}$
and finds again the equations of the one-dimensional case,
i.e.\
one only needs to apply control in the unstable direction
(see e.g. \cite{claussenthesis,ClaSchu04}).
Thus stabiliy analysis
and control schemes
of the one-dimensional case
holds also for higher-dimensional
systems provided there is only one unstable direction.
For two or more positive Ljapunov exponents
on can proceed in a similar fashion
\cite{claussenthesis,ClaSchu04}.

In OGY control, the control parameter $r_t$ is made time-dependent.
The amplitude of the feedback 
$r_t=r-r_0$
added to the control parameter $r_0$
is proportional by a constant $\varepsilon$ to the distance
$x-x^*$ from the fixed point, 
i.~e. $r=r_0 + \varepsilon (x_t-x^*)$, 
and the feedback gain
can be determined from a linearization around the fixed point,
which reads, if we neglect higher order terms,
\begin{eqnarray}
f(x_t,r_o+r_t)&=& f(x^{*},r_0)
+ (x_t-x^{*}) \cdot 
\left(\frac{\partial{} f}{\partial{} x}\right)_{x^{*},r_0}
\nonumber \\ & & 
\makebox[15em][l]{$
+ r_t \cdot
\left(\frac{\partial{} f}{\partial{} r}\right)_{x^{*},r_0}
$}
% + o(\ldots)
\nonumber 
\\
&=& 
\makebox[15em][l]{$
f(x^{*},r_0)+ \lambda (x_t-x^{*})+\mu r_t
$}
%+ o(\ldots)
\nonumber 
\\
&=& 
\makebox[15em][l]{$
f(x^{*},r_0)+ (\lambda + \mu \varepsilon)\cdot (x_t-x^{*})
$}
%+ o(\ldots)
\end{eqnarray}  
The second expression vanishes for $\varepsilon=-\lambda/\mu$,
that is, in linear approximation the system arrives 
at the fixed point at the next time step, $x_{t+1}=x^{*}$.
The uncontrolled system is assumed to be unstable in the fixed point,
% therefore we have the situation
i.\ e.\ $|\lambda|>1$.
The system with applied control is stable if the 
absolute value of the eigenvalues of the iterated map
is smaller than one,
\begin{eqnarray}
|x_{t+1}-x^{*}|=
|(\lambda + \mu \varepsilon)\cdot (x_{t}-x^{*})|
<|x_{t}-x^{*}|
\end{eqnarray}  
Therefore $\varepsilon$ has to be chosen 
between $(-1-\lambda)/\mu$ and $(+1-\lambda)/\mu$,
and this interval is of width $2/\mu$ and independent of $\lambda$,
i.e. fixed points with arbitrary $\lambda$ can be stabilized.
This property however does not survive for delayed measurement.
\cite{Cla04},
as surveyed below.

%\clearpage
\section
[Limitations of unmodified control and simple improved schemes]
{Limitations of unmodified control and simple improved control schemes}
\label{sec_ogy_etc}
In this section the limitations of unmodified control
are discussed, both for OGY control and for difference control.
For completeness, rhythmic control 
and a state space memory control are discussed in
Sections 
\ref{sec:rhythmic} and
\ref{sec:memoryx}.
% , respectively.

\subsection{Limitations of unmodified OGY control in presence of delay}
\label{sec_unmodif}

\begin{figure}[htbp]
\noindent
\raisebox{2mm}{
\begin{minipage}{56mm}
\epsfig{file=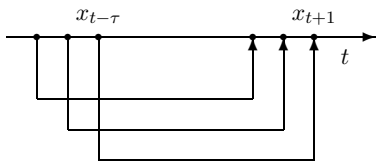,width=50mm,angle=0}
\end{minipage}
}
\hfill
\begin{minipage}{46mm}
\caption{
Unmodified control in the presence of delay (schematically).
\label{fig:cclogo1}}
\end{minipage}
\end{figure}

We want to know what limitations occur if the OGY rule
is applied without modification.
Intuitively, one expects the possibility of unstable behavour of $(\tau+1)$ control
loops that mutually overlap in the course of time 
(see Figure \ref{fig:cclogo1}).
In OGY control, the control parameter $r_t$ is time-dependent,
and without loss of generality we assume that $x^*=0$ and
that $r_t=0$ if no control is applied.
First we discuss the simplest relevant case $\tau=1$  explicitely.
For one time step delay, 
instead of $r_t = \varepsilon x_{t}$
we have the proportional feedback rule:
\begin{eqnarray}
r_t = \varepsilon x_{t-1}.
\end{eqnarray}  
Using the time-delayed coordinates $(x_t,x_{t-1})$,
the linearized dynamics of the system with
applied control
is given by
%\begin{eqnarray} 
$
\left(\begin{array}{c}x_{t+1}\\x_{t}\end{array}\right)=  \left(\begin{array}{cc}\lambda & \mu\varepsilon\\1 & 0\end{array}\right)\left( \begin{array}{c} x_{t}\\x_{t-1}\end{array}\right).
$
%\end{eqnarray}
%\begin{eqnarray} 
%\left(\begin{array}{c}x_{t+1}\\x_{t}\end{array}\right)=  \left(\begin{array}{cc}\lambda & \mu\varepsilon\\1 & 0\end{array}\right)\left( \begin{array}{c} x_{t}\\x_{t-1}\end{array}\right).
%\end{eqnarray}

\begin{figure}[htbp]
\noindent
\raisebox{2mm}{
\begin{minipage}{80mm}
\epsfig{file=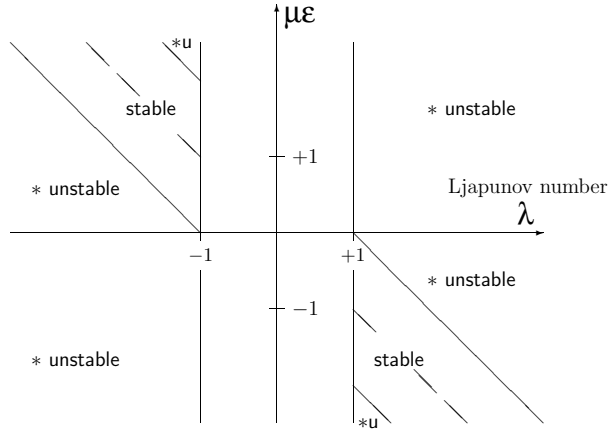,width=80mm,angle=0}
\end{minipage}
}
\hfill
\begin{minipage}{28.5mm}
\caption{
Stability range of OGY control
\label{fig:ogystab}}
\end{minipage}
\end{figure}

\noindent
The eigenvalues of 
\footnotesize
$\left(\begin{array}{cc}\lambda & \mu\varepsilon\\1 & 0\end{array}\right)$
\normalsize
are given by
$
\alpha_{1,2}=\frac{\lambda}{2}
\pm \sqrt{\frac{\lambda^2}{4}+\varepsilon\mu}.
$
Control can be achieved 
with $\varepsilon$ being
in an interval 
$]-1/\mu,(1-\lambda)/\mu[$ with the 
width $(2-\lambda)/\mu$
(see Figure \ref{fig:ogystab}).

In contrast to the not-delayed case, we have a requirement
$\lambda<2$ for the Lyapunov number: 
Direct application
of the OGY method fails for systems with a Lyapunov number 
of 2 and higher
\cite{Cla98,Cla04}.
This limitation is caused by the additional degree of freedom
introduced in the system due to the time delay.

Now we consider the general case.
If the system is measured delayed by $\tau$ steps,
$
r_t = \varepsilon x_{t-\tau},
$
we 
can
write the dynamics in time-delayed coordinates
%%$\vec{y}(t) := $
$(x_t, x_{t-1}, x_{t-2},
\ldots x_{t-\tau})^{\mbox{\scriptsize\rm{}T}}$:
\begin{eqnarray}
\left(
\begin{array}{c}
x_{t+1}\\ \vdots\\\\\\\\\\\vdots\\x_{t-\tau+1}
\end{array}\right)
= 
\left(
\begin{array}{ccccccc}
\lambda&0&\cdots&&\cdots&0&\varepsilon\mu\\
1&0&&&&&0\\
0&1&\ddots&&&&\vdots\\
\vdots&&\ddots&&&&\\  
&&&&\ddots&&\\
\vdots&&&&\ddots&0&\vdots\\
0&\cdots&&\cdots&0&1&0
\end{array}
\right)
\left(
\begin{array}{c}
x_{t}\\ \vdots\\\\\\\\\\\vdots\\x_{t-\tau}
\end{array}\right)
\nonumber
\\
\label{eq:matrixogyunmodified}
\end{eqnarray}
The characteristic polynomial is given by 
(we define rescaled coordinates $\tilde{\alpha} := \alpha/\lambda$
and $\tilde{\varepsilon} = \varepsilon\mu/\lambda^{\tau+1}$)
\\[-1ex]
\begin{eqnarray}
0&=&P(\alpha)=(\alpha-\lambda)\alpha^{\tau}-\varepsilon\mu
~~~~~~~~
\nonumber \\ 
\mbox{or}
~~~~~~~~
0&=&P(\tilde{\alpha})=(\tilde{\alpha}-1)\tilde{\alpha}^{\tau}
-\tilde{\varepsilon}.
\label{eq:charpoly}
\end{eqnarray}

%\clearpage

% \clearpage
\begin{figure}[htbp]
\begin{center}
\noindent
\vspace*{-1ex}
\epsfig{file=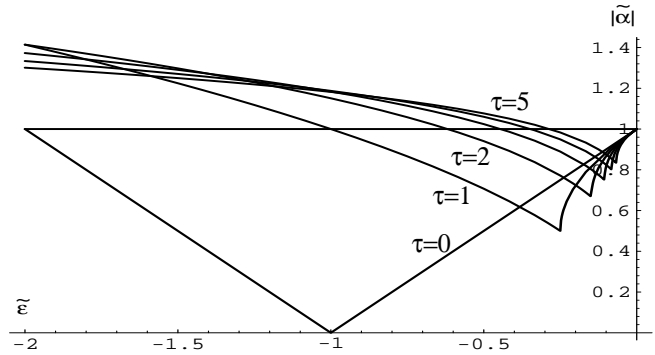, angle=0, width=0.6\columnwidth}
\vspace*{-1ex}
\end{center}
\caption{Control intervals for several time delays $\tau=0\ldots{}5$:
The plots show the maximal absolute value of the eigenvalues
as a function of the rescaled control gain $\tilde{\varepsilon}$.
Values of $|\tilde{\alpha}|=1/\lambda$ correspond to 
$|\alpha|=1$ in (\protect\ref{eq:charpoly})
without rescaling, so one can obtain the range 
$]\varepsilon_-,\varepsilon_+[$ for which control is successfully achieved.
\label{fig:kurven4}}
\end{figure}

\noindent
Fig.~\ref{fig:kurven4} shows the maximum of the
absolute value of the eigenvalues.
%%%, for $\tau=0,1,\ldots,5$.
In rescaled coordinates $\tilde{\alpha}=1/\lambda$ corresponds to
a control interval $\tilde{\varepsilon}_{\pm}(\tau,\lambda)$.
For
\begin{eqnarray}
\lambda_{\rm max} = 1 + \frac{1}{\tau}
\label{lmaxvontau}
\label{eq:lambdamaxogy}
\end{eqnarray}
the control interval vanishes, and for $\lambda\geq\lambda_{\rm max}(\tau)$
no control is possible
\cite{Cla98,Cla04}.
%
%%% Equation~(\ref{eq:lambdamaxogy}) and the subsequently derived stability diagrams are the main result of this paper and are transferred to difference control in section~\ref{sec:diffkontr}.
%
If we look at the Lyapunov exponent $\Lambda:=\ln \lambda$
instead of the Lyapunov number,
we find with $\ln x < (x-1)$ the inequality
\\[-3ex]
\begin{eqnarray}
\Lambda_{\rm max} \cdot \tau < 1.
\label{eq:controllability}
\end{eqnarray}
\\[-3ex]
Therefore, delay time and Lyapunov exponent limit each other 
if the system is to be controlled.
This is consistent with the loss of knowledge in the system
by exponential separation of trajectories.

%\clearpage
\subsection{Stability diagrams derived by the Jury criterion}
\index{Jury criterion}
For small $\tau$ one can derive easily the borders of the
stability area with help of the Jury criterion
\cite{claussenthesis,Cla04}.
%%(see Appendix~\ref{appj0}).
\label{appj0}
\noindent
The Jury criterion \cite{jury} gives a sufficient and necessary condition
that all roots of a given polynomial are of modulus smaller than 
unity.
Given a polynomial 
%\begin{eqnarray}
$
P(x)=
a_n x^n + a_{n-1} x^{n-1} + \cdots + a_1 x + a_0, 
$
%\nonumber\end{eqnarray}
one applies the iterative {\it Jury table}
scheme:
\\[-3ex]
\begin{eqnarray}
\forall_{ 0 \leq i \leq n}
\;\;\;\;
b_i &:=& a_{n-i} 
\nonumber
\\
\alpha_n &:= & b_n/a_n 
\nonumber
\\
\forall_{ 1 \leq i \leq n}
\;\;\;\;
a_{i-1}^{\rm new} 
&:=& a_i - \alpha_n b_i
\nonumber
\end{eqnarray}
\\[-3ex]
giving $\alpha_n$ and
coefficients $a_{n-1}{\ldots}a_0$ for the next iteration.
The Jury criterion states that the eigenvalues are of modulus smaller 
than unity if and only if 
%\begin{eqnarray}
$
\forall_{ 1 \leq i \leq n}
%% \;\;
|\alpha_i| < 1.
$
%\nonumber \end{eqnarray}  
%
\begin{figure}[htbp]
\noindent
\raisebox{15mm}{
\begin{minipage}{56mm}
%\begin{center}
 \epsfig{file=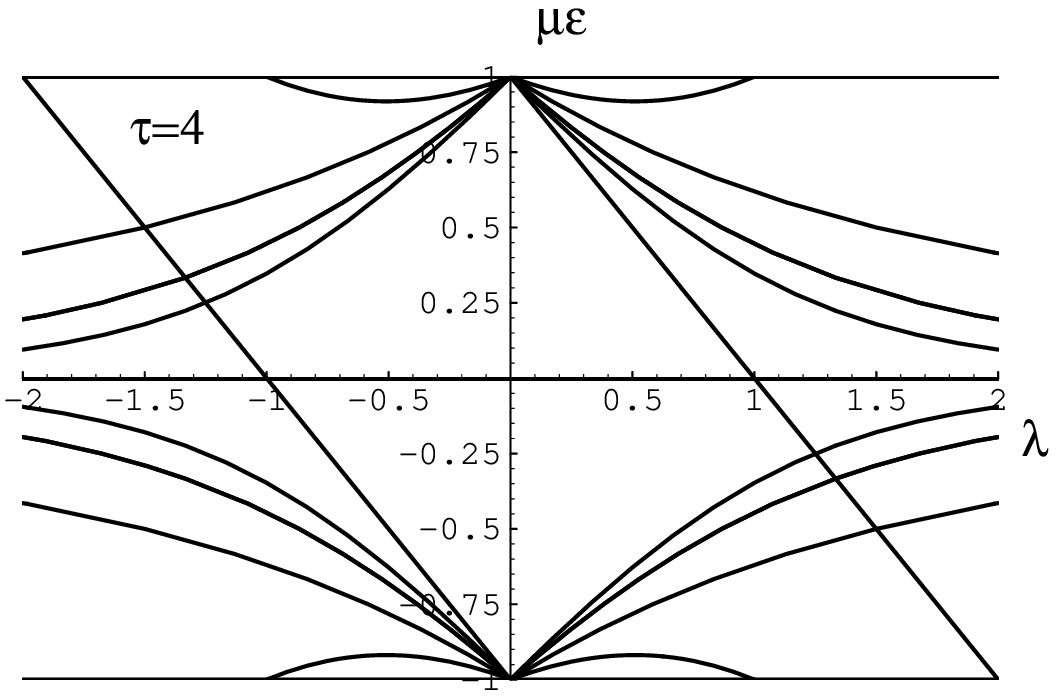, angle=270,width=55mm}
% width=55mm
%\\[-2ex]
%\end{center}
\end{minipage}
}
\hfill
\begin{minipage}{60mm}
\caption{
Complete Jury diagram
 for $\tau=4$. 
%(see Appendix~\ref{appj1}).
\label{fig:jury_tau_1}}
\end{minipage}
%\end{figure}
%\begin{figure}[htbp]
\begin{center}
\noindent
 \epsfig{file=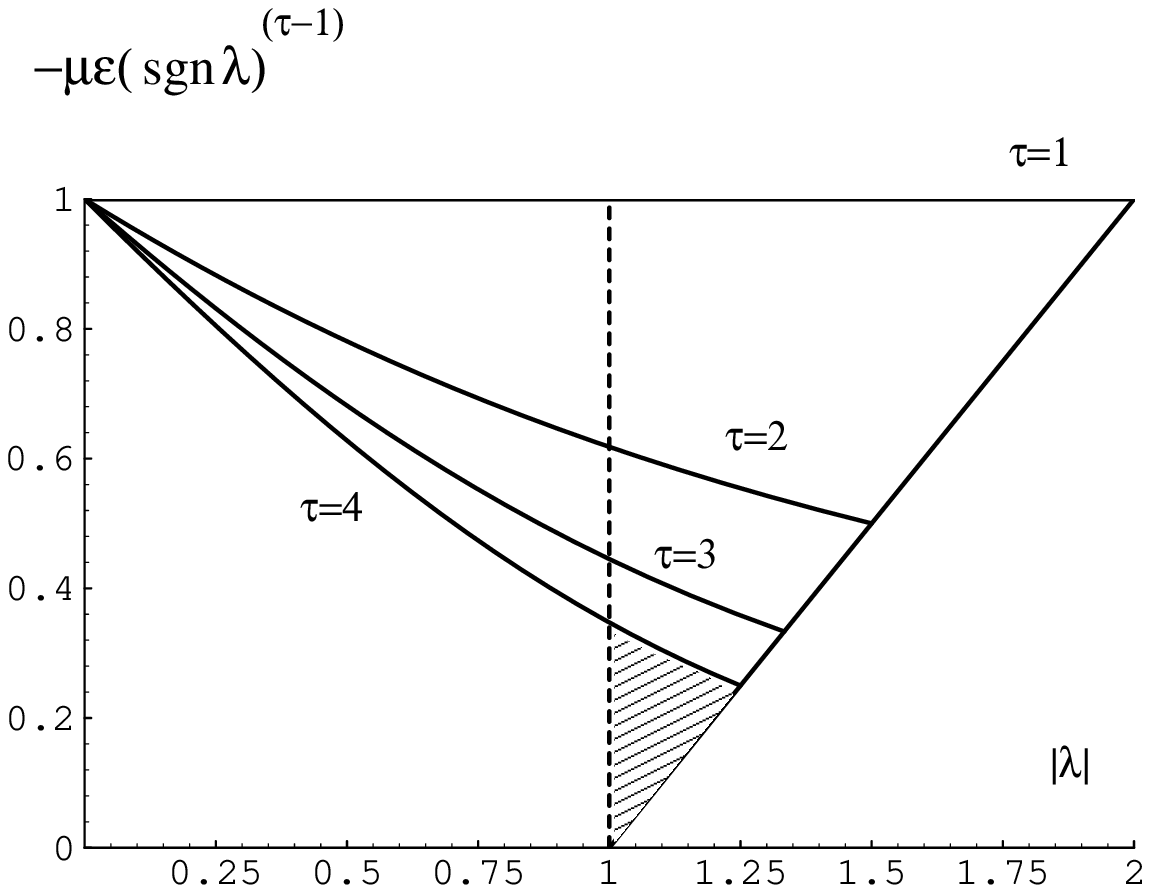, angle=0,width=0.55\columnwidth}
%0.62\columnwidth
%\\[-4ex]
\end{center}
\caption{
Stability areas for $\tau=1, 2, 3, 4$, combined.
Only for $|\lambda|>1$ control is necessary (dashed line),
and the stability area 
(shaded for $\tau=4$)
extends to 
$|\lambda_{\rm{}max}|= 2, \, 3/2, \, 4/3, \, 5/4$.
Note that still both positive and negative $\lambda$ can 
be controlled.
The abscissa $-\mu\varepsilon({\rm sgn}\lambda)^{(\tau-1)}$
takes into account that for odd $\tau$ a negative
$\mu\varepsilon$ is required, independent of the sign of $\lambda$.
\label{fig:tau1234quadrant}}
\end{figure}
\noindent
The criterion gives $2n$ (usually partly redundant)
inequalities that define hypersurfaces 
in coefficient space. 
%
%The hypersurfaces are given by algebraic equations; 
%it is not necessary to compute the roots of the polynomial.
%While the Jury criterion is extremely helpful for small $n$ and for
%numeric purposes, the hypersurface equations become unwieldy
%for large~$n$, and one has to select the relevant hypersurface equations.
%
The complete set of lines is shown in Figure~\ref{fig:jury_tau_1}
for
$\tau=4$ to illustrate the redundancy of the inequalities
generated by the Jury table.
For $\tau=1$, the Jury coefficients are given by
$\alpha_1 = -{\lambda}/{(1+\mu\varepsilon)}$ 
and $\alpha_2 = -\mu\varepsilon$.
%% and for $\tau=2$ to $\tau=4$ the  corresponding expressions are shown in Appendix~\ref{appj1}.
%
% The equations $\alpha_i=\pm{}1$ can, although the characteristic polynomial (\ref{eq:charpoly}) itself is of degree 5, be solved for one variable (giving large expressions).
%
% Only four (three for $\tau=1$) of the $2n$ inequalities constitute the border of stability, and unfortunately it seems one has to select them by hand.
%
Control is only necessary for $|\lambda|>1$, and by folding the relevant
stability area into the same quadrant 
one obtains 
Fig.~\ref{fig:tau1234quadrant}
showing how $\lambda_{\rm max}$ decreases for increasing~$\tau$.

%\epsfig{file=jury1.eps,width=50mm,angle=0}
%\epsfig{file=jury2.eps,width=50mm,angle=270}
%\epsfig{file=jury3.eps,width=50mm,angle=0}
%\epsfig{file=jury4.eps,width=50mm,angle=0}

%\clearpage

\subsection
[Stabilizing unknown fixed points: Unmodified difference control]
{Stabilizing unknown fixed points: Limitiations of unmodified difference control}
\label{sec:diffkontr}
\begin{figure}[htbp]
\noindent
\begin{center}
 \epsfig{file=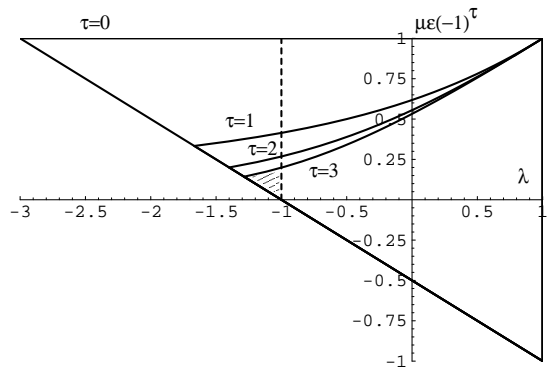, angle=0 ,width=0.5\columnwidth}
% 0.62\columnwidth
\\[-5ex]
\end{center}
\caption{
Difference feedback for $\tau=0,1,2,3$:
Stability borders derived by the Jury criterion
\cite{claussenthesis,Cla04}.
%(see Appendix~\ref{appj2}).
The stability diagram of the non-delayed case $\tau=0$ has already been given 
in \cite{bielawski93a}.
From $\lambda=-1$ (dashed line) to $\lambda=+1$ the
system ist stable without control.
For each $\tau$, 
control is effective only within the respective 
area (shaded for $\tau=3$). 
\label{fig:jurydiff01}}
\end{figure}

\index{Difference control}
As the OGY approach discussed above requires the knowledge of the 
position of the fixed point, one may wish to stabilize 
purely by feeding back differences of the system variable 
at different times.
This becomes relevant in the case of parameter drifts
\cite{Cla98} which often can occur in experimental
situations.
A time-continuous strategy 
$r(t)=\varepsilon(x(t)-x(t-\tau_{\sf d}))$
has been introduced by
Pyragas \cite{pyragas92}, 
where $r(t)$ is updated continuously and 
$\tau_{\sf d}$ matches the period of the
unstable periodic orbit.
The time-discrete counterpart
(i.e.\ control amplitudes are calculated
every Poincar\'e section)
is the difference control scheme \cite{bielawski93a}:
For control without delay, a simple difference control 
strategy 
\begin{eqnarray}
r_t  = \varepsilon (x_{t-\tau}-x_{t-\tau-1})
\end{eqnarray}
is possible for 
$\varepsilon\mu=-\lambda/3$, and eigenvalues of modulus smaller 
than unity of the matrix
 {\small $ \left( \begin{array}{cc}
 \lambda +\varepsilon\mu & - \varepsilon\mu  \\ 1 & 0
 \end{array} \right) $ \normalsize}
are obtained only for $-3 < \lambda < +1 $, so this method stabilizes 
only for oscillatory repulsive fixed points with $-3 < \lambda < -1$
\cite{bielawski93a},
see the $\tau=0$ case in Figure \ref{fig:jurydiff01}).

% \clearpage

We can proceed in a similar fashion as for OGY control.
In the presence of $\tau$ steps delay 
the linearized dynamics of
difference control
%\begin{eqnarray}
%r_t  = \varepsilon (x_{t-\tau}-x_{t-\tau-1})
%\end{eqnarray}
is given by
\begin{eqnarray}
\left(
\begin{array}{c}
x_{t+1}\\ \vdots\\\\\\\\\\\vdots\\x_{t-\tau}
\end{array}\right)
= 
\left(
\begin{array}{ccccccc}
\lambda&0&\cdots&&0&\varepsilon\mu&-\varepsilon\mu\\
1&0&&&&&0\\
0&1&\ddots&&&&\vdots\\
\vdots&&\ddots&&&&\\
&&&&\ddots&&\\
\vdots&&&&\ddots&0&\vdots\\
0&\cdots&&\cdots&0&1&0
\end{array}
\right)
\left(
\begin{array}{c}
x_{t}\\ \vdots\\\\\\\\\\\vdots\\x_{t-\tau-1}
\end{array}\right)
\nonumber
\end{eqnarray}
in delayed coordinates
$(x_t,x_{t-1},\ldots x_{t-\tau-1})$,
and the characteristic polynomial is given by
\begin{eqnarray}
0=(\alpha-\lambda)\alpha^{\tau + 1}+ (1-\alpha) \varepsilon\mu.
\end{eqnarray}

\noindent
As we have to use $x_{t-\tau-1}$  in addition to 
$x_{t-\tau}$, the system is of dimension $\tau+2$,
and the lower bound of Lyapunov numbers that can be controlled
are found to be 
\begin{eqnarray}
\lambda_{\rm{}inf} = - \frac{3+2\tau}{1+2\tau} 
=- \left(1+\frac{1}{\tau+ {1}/{2} } \right)
\label{eq:lambdamaxdiff}
\end{eqnarray}
and the asymptotic control amplitude at this point is 
% given by
\begin{eqnarray}
\varepsilon\mu = \frac{ (-1)^{\tau} }{1+2\tau}.
\end{eqnarray}

\typeout{Fig.4 here}

%\clearpage

The stability area in the $(\mu\varepsilon,\lambda)$ plane 
is bounded by the lines $\alpha_i=\pm{}1$ 
where $\alpha_i$ are the coefficients given by the 
Jury criterion \cite{jury}
 (see Figure \ref{fig:jurydiff01}).
For $\tau=0$, the Jury coefficients are
$
\alpha_1 =- \frac{\lambda+\varepsilon\mu}{1+\varepsilon\mu}
$
and
$
\alpha_2 = \varepsilon\mu.
$
For $\tau=1$ to $\tau=3$, the Jury coefficients are
given in 
\cite{Cla04}.
% Appendix~\ref{appj2}.

The controllable range is smaller than for
the unmodified OGY method, and is restricted
to oscillatory repulsive fixed points with 
$\lambda_{\rm{}inf} < \lambda \leq -1$.
Thus, delay severely reduces the number of controllable
fixed points,
and one
 has to develop special control strategies
for the control of delayed measured systems.
A striking observation is that inserting $\tau+\frac{1}{2}$ for $\tau$
in eq.~(\ref{eq:lambdamaxogy}) exactly leads to the expression in
eq.~(\ref{eq:lambdamaxdiff}) which reflects the fact that the difference
feedback control can be interpreted as a discrete first derivative, 
taken at
time $t-(\tau+\frac{1}{2})$.
Thus the controllability relation~(\ref{eq:controllability})
holds again.

As $\lambda^{-1}$ is implying a natural time scale 
(that of exponential separation) of an orbit, it is
quite natural that control becomes limited
by a border proportional to a product of $\lambda$
and a feedback delay time. 
Already  without the additional
difficulty of a measurement delay
this is expected to appear 
for any control scheme
that itself 
is using time-delayed 
feedback:
E.g.\ the extensions 
of time-discrete control schemes discussed in \cite{socolar98}
with an inherent Lyapunov number limitation
due to memory terms,
and the experimentally widely applied 
time-continuous schemes 
Pyragas and ETDAS
\cite{just97,just99pla,franc99}.
Here Pyragas control has 
the Lyapunov exponent limitation $\Lambda\tau_{\sf p}\leq{}2$ 
together with the requirement of the Floquet multiplier 
of the uncontrolled orbit having an imaginary part of
$\pi$, meaning that deviations from the orbit after one
period experience to be flipped around the orbit by that angle,
which is quite the generic case \cite{just99pre}.
This nicely corresponds with the requirement
of a negative Lyapunov number that appears in difference control.
A positive Lyapunov number in the time-discrete picture
corresponds to a zero flip of the time-continuous orbit,
and is consistently uncontrollable in both schemes.

Recently, the influence of a control loop latency 
has also been studied for continuous time-delayed feedback
\cite{just99pre} by Floquet analysis,
obtaining a critical value
%  of $\tau(1-\lambda \tau_{\sf p}/2)/(\lambda \tau_{\sf p})$ 
for the measurement delay $\tau$, 
that corresponds to a maximal Lyapunov exponent
$\log|\lambda_{\sf inf}|= \Lambda\tau_{\sf p} = \frac{1}{1/2 + \tau/\tau_{\sf p}}$,
where $\tau_{\sf p}$ is the orbit length and matched feedback delay.
By the log inequality that again translates (for small Ljapunov exponents)
to our result for the time-discrete
difference control.
An exact coincidence could not be expected, as 
in Pyragas control the feedback difference is computed
continuously sliding with the motion along the orbit,
where in difference control it is evaluated within each
Poincar\'e section.
For the ETDAS scheme with latency, a detailed analysis is performed in
\cite{hovel}, showing that the range of stability can be extended  
compared to the Pyragas scheme.
Although the time-continuous case (as an a priori infinite-dimensional
delay-differential system) could exhibit much more
complex behaviour, it is, however astonishing that for all three methods,
OGY, difference, and Pyragas control,
the influence of measurement delay mainly results in the same
limitation of the controllable Lyapunov number.

%\clearpage
\subsection{Rhythmic control schemes: Rhythmic OGY control}
\label{sec:rhythmic}
\index{Rhythmic control}
\begin{figure}[htbp]
\noindent
\raisebox{2mm}{
\begin{minipage}{56mm}
\epsfig{file=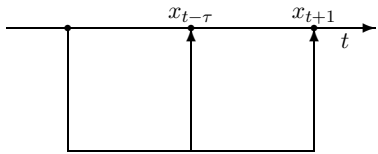,width=50mm,angle=0}
\end{minipage}
}
\hfill
\begin{minipage}{57mm}
\caption{
Rhythmic control (schematically).
Keeping control quiet for $\tau$ intermediate time steps
avoids the additional degrees of freedom. 
However, the effective Lyapunov number to be controlled
then is raised to $\lambda^{\tau+1}$.
\label{fig:cclogo2}}
\end{minipage}
\end{figure}

%\subsection{Rhythmic OGY control}
\label{sec:mem}
As pointed out for difference control 
in the case $\tau=0$ in 
\cite{bielawski93a},
%,claussenthesis,Cla03}
one can eliminate the additional degrees of freedom 
caused by the delay term. 
One can 
restrict himself to 
apply control rhythmically only every $\tau+1$ timesteps
($\tau+2$ for difference control), and then leave the system uncontrolled
for the remaining timesteps (see Fig.~\ref{fig:cclogo2}).
Then $\varepsilon=\varepsilon(t)$ 
 \clearpage\noindent
appears to be time-dependent with
\begin{eqnarray}
\varepsilon ({t {\rm ~  mod ~ } \tau}) = (\varepsilon_0,0,\ldots,0)
\end{eqnarray}
and, after $(\tau+1)$ iterations of (\ref{eq:matrixogyunmodified}), 
we again have a matrix as in (\ref{eq:matrixogyunmodified}),
but with $\lambda^{\tau+1}$ instead of $\lambda$.
Equivalently, we can write 
\begin{eqnarray}
x_{t+(\tau+1)}=\lambda^{\tau+1}x_t+\varepsilon_0\mu x_t.
\end{eqnarray}
What we have done here, is: controlling the $(\tau+1)$-fold
iterate of the original system.
This appears to be formally elegant, but leads to
practically uncontrollable high effective
Lyapunov numbers $\lambda^{\tau+1}$ for both large $\lambda$
and large~$\tau$.

Even if the rhythmic control method is of striking simplicity,
it remains unsatisfying that control is kept quiet,
or inactive, for $\tau$ time steps. 
Even if the state of the system $x$ is known delayed by
$\tau$, one knows (in principle) the values of $x_t$ for
$t<\tau$, and one could (in principle) store the values
$\delta{}r_{t-\tau}\ldots{}\delta{}r_{t}$ of the control
amplitudes applied to the system. 
This can be done, depending on the timescale, by analog 
or digital delay lines, or by storing the values in a computer
or signal processor 
(observe that
there are some 
intermediate
frequency ranges where an
experimental setup is difficult).

Both methods, rhythmic control and simple feedback 
control in every time step, have their disadvantages:
For rhythmic control it is necessary to use rather large 
control amplitudes, in average $\lambda^\tau / \tau$, 
and noise sums up to an amplitude increased by factor $\sqrt{\tau}$.
For simple feedback control the dimension of the system
is increased and the maximal controllable Lyapunov number
is bounded by (\ref{lmaxvontau}).
One might wonder if there are control strategies 
that avoid these limitations.
This has necessarily to be done by applying control in each time step,
but with using knowledge what control has been applied 
between the last measured time step $t-\tau$ and~$t$.
This concept can be implemented in at least two ways,
by storing previous values of $x_t$
(Section \ref{sec:memoryx})
 or previous values of 
$\delta r_t$ (LPLC, Section \ref{sec:lplc}
and MDC, Section \ref{sec:mdc}).

\subsection{Rhythmic difference control}
\begin{figure}[htbp]
\noindent
\epsfig{file=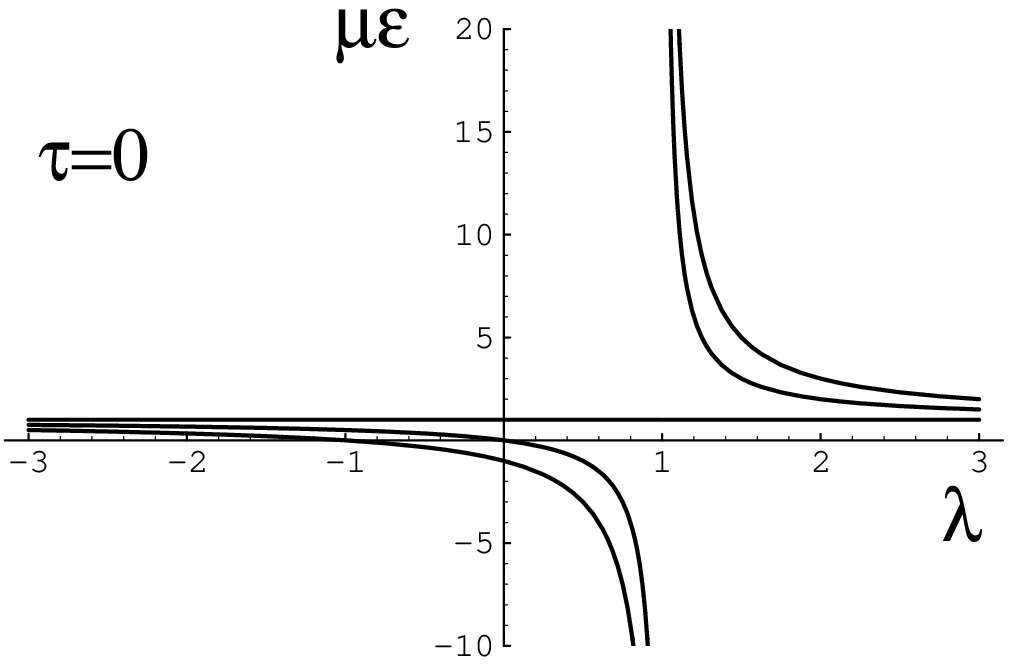,width=.31\columnwidth,angle=270}
\epsfig{file=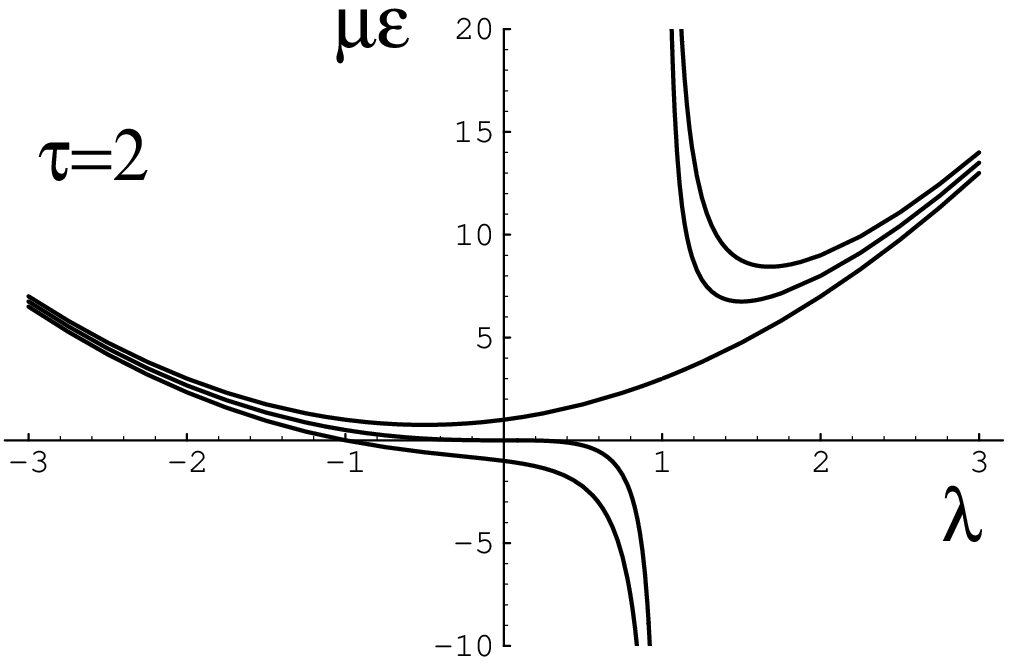,width=.31\columnwidth,angle=270}
\epsfig{file=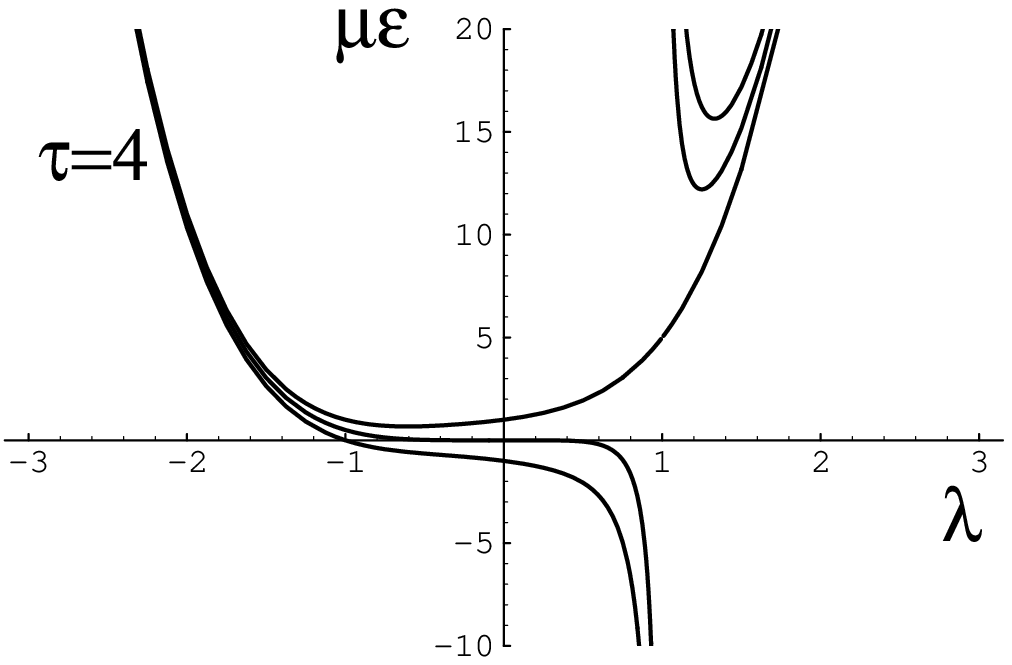,width=.31\columnwidth,angle=270}
\\
\epsfig{file=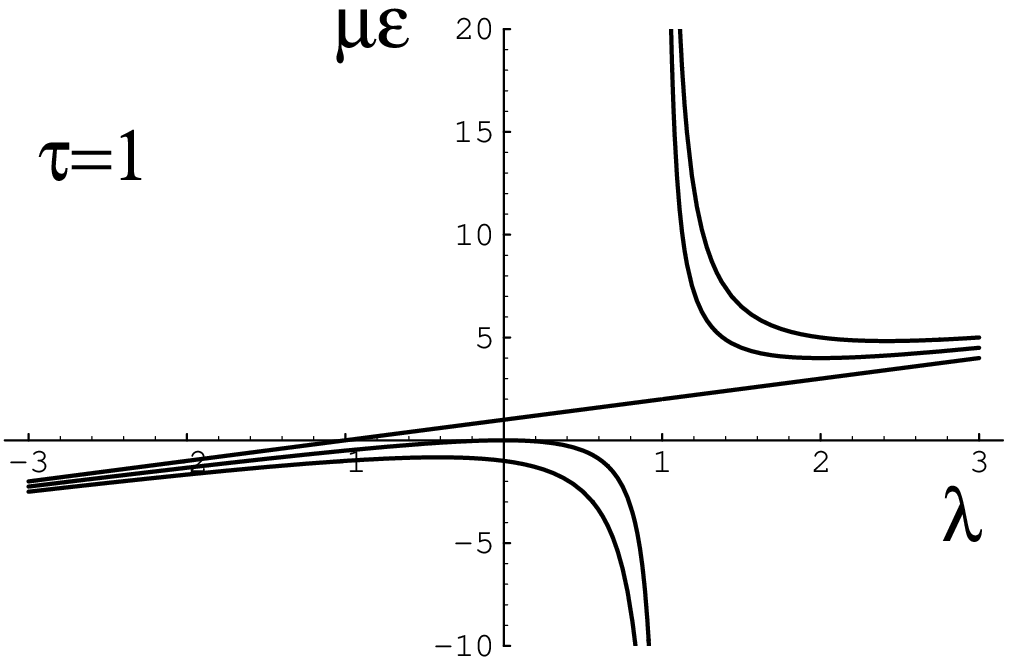,width=.31\columnwidth,angle=270}
\epsfig{file=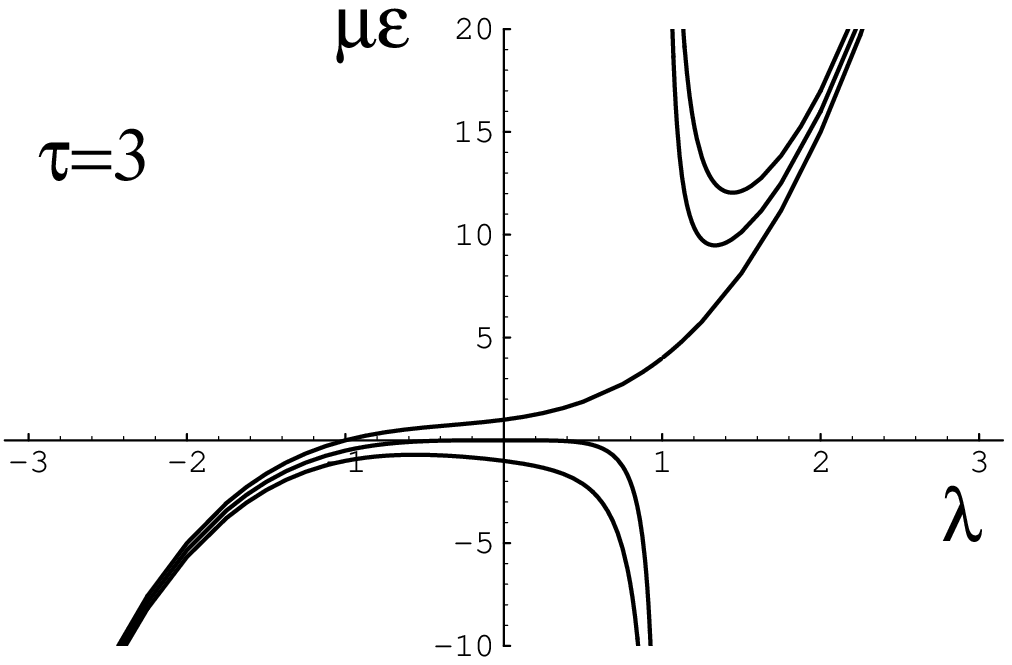,width=.31\columnwidth,angle=270}
\epsfig{file=diff_tau5.eps,width=.31\columnwidth,angle=270}
%\begin{center}\begin{tabular}{cc}
%\epsfig{file=diff_tau0.eps,width=47mm,angle=270} &
%\epsfig{file=diff_tau1.eps,width=47mm,angle=270}\\
%\epsfig{file=diff_tau2.eps,width=47mm,angle=270}&
%\epsfig{file=diff_tau3.eps,width=47mm,angle=270}\\
%\epsfig{file=diff_tau4.eps,width=47mm,angle=270}&
%\epsfig{file=diff_tau5.eps,width=57mm,angle=270}
%\end{tabular}\end{center}
\caption{Stability area of rhythmic difference control
for 
$\tau=0,1,2,3,4,5$.
}
 \label{stab_rdc012}
\label{fig:peri_difftau012345}
\end{figure}

\index{Rhythmic difference control}
To enlarge the range of controllable $\lambda$, one
again has the possibility 
to reduce the dimension of the
control process in linear approximation
to one by applying control every $\tau+2$ time steps.
\begin{eqnarray}
x_{t+1}
&=& 
\lambda x_t +
\mu \varepsilon (x_{t-\tau}-x_{t-\tau-1})
\label{eq:peridiff_opt}
\\
&=&
(\lambda^{\tau+1} +\mu \varepsilon \lambda - \mu \varepsilon)
x_{t-\tau-1}
\nonumber
\end{eqnarray}
and the goal $x_{t+1} \stackrel{!}{=} 0$ can be fulfilled by
\begin{eqnarray}
\mu \varepsilon =
- \frac{\lambda^{\tau+1}}{1-\lambda}
\end{eqnarray}
 \clearpage   
 \noindent
One has to choose $\mu \varepsilon$ between 
$\mu \varepsilon_{\pm}= - \frac{\lambda^{\tau+1} \pm 1}{1-\lambda} $
to achieve control as shown in Fig.~\ref{fig:peri_difftau012345}.
The case $\tau=0$ has already been discussed in
\cite{bielawski93a}.
%%%,claussenthesis,Cla03,schusterstemmler}.
With rhythmic control,
there is no range limit for $\lambda$, and even fixed points
with positive $\lambda$ can be stabilized by this method.
%\clearpage

\noindent
When using differences for periodic feedback, one
still has the problem that the control gain increases by $\lambda^\tau$,
and noise sums up for $\tau+1$ time steps before the next control
signal is applied.
Additionally, now there is a singularity for $\lambda=+1$ in the ``optimal''
control gain given by (\ref{eq:peridiff_opt}).
This concerns fixed points where differences $x_t-x_{t-1}$ 
when escaping from the fixed point are naturally
small due to a $\lambda$ near to $+1$.

%\clearpage

Here one has to decide between using a large control gain 
(but magnifying noise and finite precision effects)
or using a small control gain of order 
$\mu\varepsilon_{-}(\lambda=+1) =\tau+1$ 
(but having larger eigenvalues and therefore slow convergence).

%\clearpage
Two other strategies that have been discussed by Socolar and Gauthier
\cite{socolar98} are discretized versions of time-continuous methods.
Control between $\lambda= -(3+R)/(1-R)$  and  $\lambda= -1$ is possible with
discrete-ETDAS ($R<1$)
%\begin{eqnarray}
$
r_t=\varepsilon \sum_{k=0}^\infty R^k (x_{t-k}-x_{t-k-1})
$
%\end{eqnarray}
and control between $\lambda= -(N+1)$ and  $\lambda= -1$ 
is acheived with discrete-NTDAS (let $N$ be a positive integer)
which is
defined by 
%\begin{eqnarray}
$
r_t=\varepsilon \left(x_t-\frac{1}{N}\sum_{k=0}^N x_{t-k}\right).
$
%\end{eqnarray}
Both methods can be considered to be of advantage even in 
time-discrete control in the Poincar\'e section, e.g. if the
number of adjustable parameters has to be kept small.
Whereas these methods are mainly applied in time-continuous 
control, especially in analogue or optical experiments,
for time-discrete control the MDC strategy described below
allows to overcome 
the
% limitations
% in the
 Lyapunov number
limitations.

\clearpage
\subsection{\mbox{A simple memory control scheme: Using state space memory}}
\label{sec:memoryx}
\index{State space memory control}
\index{Multiple delay lines}
\begin{figure}[htbp]
\noindent
\raisebox{2mm}{
\begin{minipage}{56mm}
\epsfig{file=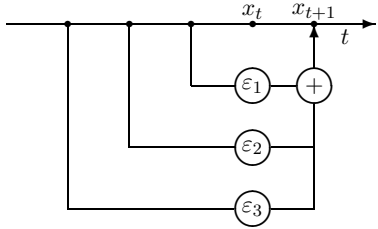,width=50mm,angle=0}
\end{minipage}
}
\hfill
\begin{minipage}{54mm}
\caption{
A state space memory control (schematically).
For electronic or optic analogue circuits, the idea to use additional delay
lines is appealing, though the applicability is restricted
to the $\tau=1$ OGY case
(which will cover most experiments).
\label{fig:cclogo3}}
\end{minipage}
\end{figure}
We extend the single delay line by 
several artificial delay lines (see Fig.~\ref{fig:cclogo3}), 
each with an externally tuneable control gain coefficient
\cite{claussenthesis,ClaSchu04}:
\begin{eqnarray}
r_t &=& \varepsilon_1  x_{t-1}
 + \varepsilon_2  x_{t-2} 
 + \ldots 
+  \varepsilon_{n+1}  x_{t-n-1} 
\end{eqnarray}
%
%\clearpage
%
For $n$ steps memory (and one step delay) the control matrix is 
\begin{eqnarray}  
\!\!
\left(
\!\!
\begin{array}{c}
x_{t+1}\\ \vdots\\\\\\\\\\\vdots\\x_{t-n}
\end{array}
\!\!
\right)
\!\!
= 
\!\!
\left(
\!\!
\begin{array}{ccccccc}
\lambda& \varepsilon_1
&\cdots&& &\varepsilon_{n}& \varepsilon_{n+1}\\
1&0&&&&&0\\
0&1&\ddots&&&&\vdots\\
\vdots&&\ddots&&&&\\  
&&&&\ddots&&\\
\vdots&&&&\ddots&0&\vdots\\
0&\cdots&&\cdots&0&1&0
\end{array}
\!\!
\right)
\!\!
\left(
\!\!
\begin{array}{c}
x_{t}\\ \vdots\\\\\\\\\\\vdots\\x_{t-n-1}
\end{array}
\!\!
\right)
\end{eqnarray}
%   \clearpage
with the
characteristic
polynomial
%\begin{eqnarray}
$
(\alpha-\lambda)\alpha^{n+1} + \sum_{i=1}^{n} \varepsilon_i \alpha^{n-i}.
$
%\end{eqnarray}
% \clearpage
We can choose $\alpha_1 = \alpha_2 = \ldots \alpha_{n+2} = -\lambda/(n+2)$
and evaluate optimal values for all $\varepsilon_i$ 
by comparing with the coefficients of the product
$\prod_{i=1}^{n+2} (\alpha-\alpha_i)$.
This method allows control up to $\lambda_{\rm{}max}=2+n$,
%therefore
thus
 arbitrary $\lambda$ can be controlled if a memory
length of $n>\lambda-2$ and the optimal coefficents $\varepsilon_i$
are used.

For more than one step delay, one has the situation
$\varepsilon_1=0,\ldots,\varepsilon_{\tau-1}=0$.
This prohibits the 'trivial pole placement' given above,
(choosing all $\alpha_i$ to the same value) 
and therefore reduces the maximal controllable $\lambda$
and no general scheme for optimal selection of the $\varepsilon_i$ applies.
One can alternatively use the LPLC method described below,
%%%%%%%% new paragraph start
which provides an optimal control scheme.
One could wonder why to consider the previous state memory scheme at all
when it does not allow to make all eigenvalues zero in {\sl any} case.
First, the case of up to one orbit delay and moderately small
$\lambda$ already covers many low-period orbits.
Second, there may be experimental setups where the feedback of 
previous states through additional delay elements and 
an analog circuit is experimentally more feasible
than feedback of past control amplitudes.

%In summary, all three methods discussed in this section have principal
%restrictions, but may be of advantage in special situations especially
%when a simple setup is required.

%
\clearpage
\section{Optimal Improved Control schemes}
\vspace*{-1ex}
\subsection{Linear predictive logging control (LPLC)} \label{sec:lplc}
\begin{figure}[htbp]
\noindent
\raisebox{2mm}{
\begin{minipage}{56mm}
\epsfig{file=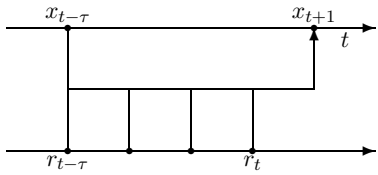,width=50mm,angle=0}
\end{minipage}
}
\hfill
\begin{minipage}{57mm}
\caption{
Linear predictive logging control (schematically).
In LPLC, all intermediately applied control amplitudes are employed 
for a linear prediction. 
A corresponding scheme (MDC, Section \ref{sec:mdc}) exists also for 
difference control.
\label{fig:cclogo4}}
\end{minipage}
\end{figure}

\index{Linear predictive logging control}
\index{LPLC}
If it is possible to store the previously applied control amplitudes
$r_t, r_{t-1},\ldots$, then one can predict the actual state $x_t$ of 
the system using the linear approximation around the fixed point
(see Fig.~\ref{fig:cclogo4}).
That is, from the last measured value $x_{t-\tau}$ 
and the control amplitudes 
we compute estimated values iteratively by
\begin{eqnarray}
y_{t-i+1}  =  \lambda x_{t-i} + \mu  r_{t-i} 
\end{eqnarray}
leading to a {\it predicted} value $y_t$ of the actual system state.
Then the original OGY formula can be applied, i.~e.
$r_t= -\lambda/\mu y_t$.
In this method the gain parameters are again linear in $x_{t-\tau}$ and
all $\{ r_{t'}\}$ with $t-\tau \leq t' \leq t$, and the optimal
gain parameters can be expressed in terms of $\lambda$ and $\mu$.
\\[-7mm]

In contrast to the memory method presented in the previous subsection,
the LPLC method directs the system (in linear approximation) in one
time step onto the fixed point.
However, when this control algorithm is switched on, one has no control
applied between $t-\tau$ and $t-1$, so the trajectory has to be fairly
near to the orbit (in an interval with a length 
of order $\delta/\lambda^\tau$, where $\delta$ is the 
interval halfwidth where control is switched on).
Therefore the time one has to wait until the control can be successfully
activated is of order $\lambda^{\tau-1}$ larger than in the case of
undelayed control.

The LPLC method can also be derived as a general linear feedback
in the last measured system state and all applied control
amplitudes since the system was measured
-- by choosing the feedback
gain parameters 
in a way that the linearized system has all eigenvalues 
zero.
The linear ansatz
\begin{eqnarray}
r_{t} = 
\varepsilon \cdot
x_{t-\tau-i} 
+ \eta_1  r_{t-1} + \ldots  \eta_{\tau}  r_{t-\tau} 
\end{eqnarray}
leads to the dynamics in
combined delayed coordinates
\\
$(x_t, x_{t-1}, \ldots ,x_{t-\tau}, r_{t-1},  \ldots , r_{t-\tau})$
   \clearpage
\footnotesize
\begin{eqnarray}
\!\!
\left(
\!\!
\begin{array}{c}
x_{t+1}\\ 
x_t\\ ~\\~ \\[-.35ex]  \vdots \\[-.35ex] ~ \\ ~ \\ x_{t-\tau+1}\\
 r_{t} \\[3.6ex]  \vdots \\[3.6ex]  r_{t-\tau+1}
\end{array}
\!\! \right) \!\! = \!\! \left( \!\!
\begin{array}{ccccccccccc}
\lambda&0&\cdots&\cdots&0&\varepsilon&  \eta_1&\eta_2&\cdots&\cdots&\eta_{\tau} \\
 & 0 & & & & & & & & & \\
 & 1 & \ddots & & & & & & & & \\
 & & \ddots&\ddots & & & & & & & \\  
 & & & \ddots&\ddots & & & & & & \\  
 & & & & 1& 0& & & & & \\
0 &0&\cdots&\cdots&0&\varepsilon& \eta_1&\eta_2&\cdots&\cdots&\eta_{\tau} \\
 & & & & & &1 &0 & & & \\  
 & & & & & & &\ddots &\ddots & & \\  
 & & & & & & & & \ddots&\ddots & \\  
 & & & & & & & & &1 &0 
\end{array}
\!\! \right) \!\! \left( \!\!
\begin{array}{c}
x_t\\ 
x_{t-1}\\ ~\\~ \\[-.35ex] \vdots \\ ~\\ ~ \\[-.35ex]  x_{t-\tau}\\
 r_{t-1} \\[3.6ex]  \vdots \\[3.6ex] r_{t-\tau}
\end{array}
\!\!
\right)
\nonumber
\end{eqnarray}
\\[-4mm]
%\begin{eqnarray}   \end{eqnarray}
\normalsize
%
%
%
%\clearpage
\noindent
giving the characteristic polynomial
\\[-8mm]
%\clearpage
\begin{eqnarray}
0 &=& -\alpha^\tau
(
\alpha^{\tau+1}
+\alpha^{\tau} (-\lambda - \eta_1)
\nonumber\\
\nonumber
& &
\mbox{~~~~~}
+\alpha^{\tau-1} (\lambda \cdot \eta_1 - \eta_2)
+\alpha^{\tau-2} (\lambda \cdot \eta_2 - \eta_3) 
\\
&& \mbox{~~~~~} \ldots
+\alpha^{1} (\lambda \cdot \eta_{\tau-1} - \eta_{\tau}) 
+ (\lambda \cdot \eta_{\tau} -\varepsilon)   ).
\end{eqnarray}
\\[-7.5mm]
All eigenvalues can be set to zero using
$\varepsilon=-\lambda^{\tau+1}$
and $\eta_i=-\lambda^{i}$.
A generalization to
 more than one positive
Lyapunov exponent 
is given in 
\cite{ClaSchu04}.
\\[-9mm]

%\clearpage
\subsection{Nonlinear predictive logging control}
\index{Nonlinear predictive logging control}
One can also consider a 
nonlinear predictive logging control (NLPLC)
strategy
\cite{ClaSchu04} 
as the straightforward extension to the LPLC method for nonlinear
prediction. If the system has a delay of several time steps, 
the interval where control is achieved becomes too small.
However, if it is possible to extract the first nonlinearities from the 
time series, prediction (and control) can be fundamentally improved. 
In NLPLC, the behaviour of the system is predicted each time step by
a truncated Taylor series
\\[-6mm]
\begin{eqnarray}  
x_{t+1}  =  \lambda x_t +\frac{\lambda_2}{2} x_t^2 
 + \mu r_t + \frac{\mu_2}{2} r_t^2 
\nonumber 
+ \nu x_t r_t
+ o(x_t^3,
x_t^1 r_t, x_t r_t^2,r_t^3)
%,\ldots) 
\end{eqnarray}
\\[-6mm]
% with 
using
% the
 applied control amplitudes $\{r_t\}$ for each time step.
This
equation
%eqn.\
 has to be solved for $r_t$ using $x_{t+1} \stackrel{!}{=}0$.
A similar nonlinear prediction method has been described by 
Petrov and Showalter \cite{showalter96}.
They approximate the $x_{t+1}(x_t,r_t)$ surface 
directly from the time series and use it to direct the system
to any desired point.
Both Taylor approximation or Petrov and Showalter method 
can be used here iteratively, provided one knows the delay length.
Both approaches could be regarded as a nonlinear method of model
predictive control \cite{mpc},
applied to the Poincar\'e iteration dynamics.

From a practical point of view, 
it has to be mentioned that one has to know the fixed point $x^*$
more accurate than in the linear case.
Otherwise one experiences a smaller range of stability
and additionally a permanent nonvanishing control
amplitude will remain.
This may be of disadvantage especially
if the fixed point drifts in time (e.g. by other external parameters
such as temperature) or if the time series used to determine the 
parameters is too short.

\clearpage
\subsection[Stabilization of unknown fixed points: MDC
]{Stabilization of unknown fixed points: Memory difference control (MDC)} 
\label{sec:mdc}
\index{Memory difference control}
\index{MDC}
As all methods mentioned above require the knowledge of the 
position of the fixed point, one may wish to stabilize 
purely by feeding back differences of the system variable 
at different times.
Without delay, difference feedback can be used successfully for 
$\varepsilon\mu=-\lambda/3$, and eigenvalues of modulus smaller 
than unity of the matrix
{\small
$
\left(
\begin{array}{cc}
\lambda +\varepsilon\mu & - \varepsilon\mu  \\ 1 & 0
\end{array}
\right)
$
\normalsize}
are obtained only for $-3 < \lambda < +1 $, so this method stabilizes 
only for oscillatory repulsive fixed points with $-3 < \lambda < -1$
\cite{bielawski93a}.

Due to the inherent additional period one delay of difference control and MDC,
the $\tau$ period delay case of
MDC corresponds,
in terms of the number of degrees of freedom,
to the  $\tau+1$ period delay
case of LPLC.

%\clearpage
%
One may wish to generalize the linear predictive feedback
to difference feedback. 
Here, caution is advised.
In contrary to the LPLC case, 
the reconstruction of the state $x_{t-\tau}$ from 
differences $x_{t-\tau-i} - x_{t-\tau-i-1}$ 
and applied control amplitudes $r_{t-j}$ is no longer unique.
As a consequence, there are infinitely many ways to compute an estimate
for the present state of the system,
but only a subset of these leads to a controller design ensuring convergence
to the fixed point.
%
%Under
Among
 these there exists an optimal every-step control for difference
feedback with minimal eigenvalues and in this sense optimal stability.

To derive the feedback rule for MDC
\cite{Cla98,ClaSchu04,claussenthesis}, 
we directly make the linear ansatz
\begin{eqnarray}
\nonumber
r_{t} = 
\varepsilon \cdot
(x_{t-\tau-i} - x_{t-\tau-i-1})
+ \eta_1  r_{t-1} + \ldots  \eta_{\tau}  r_{t-\tau} 
\nonumber
\end{eqnarray}
%leading to
%yielding
with
 the dynamics in
combined delayed coordinates
\footnotesize
\begin{eqnarray} 
%%%{\bf M}
\left(\!\!\begin{array}{c}
x_{t+1}\\ x_t\\ ~ \\  \vdots \\[2ex]  ~ 
\\x_{t-\tau+2}\\x_{t-\tau+1}\\ 
r_{t}\\r_{t-1}\\ ~\\  \vdots\\~ \\
r_{t-\tau+1}\end{array}\!\!\right)
\!\! 
= 
\!\!
\left(
\!\!
\begin{array}{ccccccccccc}
\lambda&0&\cdots&0 &\varepsilon & -\varepsilon&  
\eta_1& \eta_2-\eta_1&\cdots&\cdots&\eta_{\tau}-\eta_{\tau+1} \\
1 & 0 & & & & & & & & & \\
 & 1 & \ddots & & & & & & & & \\
 & & \ddots&\ddots & & & & & & & \\  
 & & & \ddots&\ddots & & & & & & \\  
 & & & & 1& 0& & & & & \\
0&0&\cdots&0 &\varepsilon & -\varepsilon&  
\eta_1& \eta_2-\eta_1&\cdots&\cdots&\eta_{\tau} -\eta_{\tau+1} \\
 & & & & & &1 &0 & & & \\  
 & & & & & & &\ddots &\ddots & & \\  
 & & & & & & & & \ddots&\ddots & \\  
 & & & & & & & & &1 &0 
\end{array}
\!\!
\right)
\!\!
\left(\!\!\begin{array}{c}
x_{t}\\ x_{t-1}\\~\\ \vdots\\[2ex] ~
\\x_{t-\tau+1}\\x_{t-\tau}\\
 r_{t-1}\\ r_{t-2}\\~\\   \vdots \\~ \\ r_{t-\tau}\end{array}\!\!\right)
\nonumber
\end{eqnarray}
\normalsize
\clearpage
\noindent
giving the characteristic polynomial
%\clearpage
\begin{eqnarray}
0 &=& -\alpha^\tau
(
\alpha^{\tau+1}
+\alpha^{\tau} (-\lambda - \eta_1)
\nonumber
\\ &&
+\alpha^{\tau-1} (\lambda \cdot \eta_1 - \eta_2)
+\alpha^{\tau-2} (\lambda \cdot \eta_2 - \eta_3) 
\nonumber
\\ &&
 \ldots
+\alpha^{2} (\lambda \cdot \eta_{\tau-2} - \eta_{\tau-1}) 
\nonumber
\\ &&
+\alpha^{1} (\lambda \cdot \eta_{\tau-1} - \eta_{\tau}- \varepsilon) 
+ (\lambda \cdot \eta_{\tau} +\varepsilon)   ).
\end{eqnarray}
All eigenvalues can be set to zero using
$\varepsilon=-\lambda^{\tau+1}/(\lambda-1)$,
$\eta_{\tau}= + \lambda^{\tau}/(\lambda-1)$
and $\eta_i=-\lambda^{i}$ for $1\leq i \leq \tau-1$.
This defines the MDC method.
%\\
%The general formulas even f
For more than one positive 
Lyapunov exponent
% or  multiparameter control are given in Appendix~\ref{app_diff}.
see \cite{claussenthesis,ClaSchu04}.

%%This memory difference control (MDC) method has been demonstrated in an electronic experiment \cite{Cla98} and a plasma diode \cite{mausbachpp99}.

%\clearpage
\section{Summary}
Delayed measurement is a generic problem that can appear in
controlling chaos experiments.
In some situations it may be technically impossible to extend the
control method, then one wants to know the stability borders
with minimal knowledge of the system.

We have shown that both OGY control and difference control cannot
control orbits with an arbitrary Lyapunov number if there is
only delayed knowledge of the system.
The maximal Lyapunov number 
up to which an instable orbit can be controlled
is given by $1+\frac{1}{\tau}$ for 
OGY control and 
%$1+{1}/{(\tau+{1}/{2})}$ 
$1+\frac{1}{\tau+{1}/{2}}$
for difference control.
For small $\tau$ the stability borders can be derived by the
Jury criterion, so that the range of values for
the control gain $\varepsilon$ can be determined from the
knowledge of the Taylor coefficients $\lambda$ and $\mu$.
If one wants to overcome these limitations, 
one has to modify the control strategy.

We have presented methods to improve 
Poincar\'e-section based chaos control for 
delayed measurement. For both classes of algorithms,
OGY control and difference control, delay 
affects control, and improved control strategies
have to be applied.
Improved strategies contain one of the following 
principle ideas: Rhythmic control, control with memory
for previous states, or control with memory
for previously applied control amplitudes.
In special cases the unmodified control, 
previous state memory control, or rhythmic control
methods could be considered, especially
when experimental conditions restrict the possibilities 
of designing the control strategy.

In general, the LPLC and MDC strategies allow a so-called
deadbeat control with all eigenvalues zero; and they are
in this sense optimal control methods.
All parameters needed for controller design can be 
calculated from linearization parameters that can be
fitted directly from experimental data.
This approach has also been sucessfully 
applied in an electronic
\cite{Cla98}
and plasma 
\cite{mausbachpp99}
experiment.

%\clearpage
%\bibliographystyle{enoc}

%
\typeout{}
\typeout{}
\typeout{------------------------------------------------------------------------------}
\typeout{}
\typeout{CLA: Kommentare zur .tex Datei (Claussen, Poincare-based control...):}
\typeout{}
\typeout{CLA: Die Verwendung von caption{} generiert: Table x.y; Ursache: mir unklar.}
\typeout{}
\typeout{CLA: In den Matrizen vor Formeln (x.20) und (x.21) sind mittels z.B. [2.5ex]}
\typeout{CLA: von Hand Abstaende einjustiert, damit die erste r_t Zeile in den}
\typeout{CLA: Vektoren mit der zweiten vollstaendigen Matrixzeile (0,0,...)}
\typeout{CLA: auf gleiche Hoehe kommt. Falls am Stylefile oder Parametern}
\typeout{CLA: geaendert werden sollte, bitte im Auge behalten...}
\typeout{}
\typeout{------------------------------------------------------------------------------}
\typeout{}
\typeout{}


\begin{thebibliography}{22}


%\small

\bibitem%
%[Bielawski et al.\ 1993]
{bielawski93a}
%S.~Bielawski, D.~Derozier, and P.~Glorieux
Bielawski S.,  Derozier D.\ and Glorieux P.\
(1993).
%Experimental characterization of unstable periodic orbits by controlling c
{\sl Phys. Rev. A} {\bf 47} (4),  
2492--2495.
%\\[-7mm]

\bibitem%
%[Blakely et al.\ 2004]
{blakely} 
%{J. N. Blakely}, {L. Illing}, and {D. J. Gauthier},
Blakely J.\ N.,  Illing L.\ and  Gauthier D.\ J.\
(2004).
 Controlling Fast Chaos in Delay Dynamical Systems,
{\sl Phys. Rev. Lett.} {\bf 92},
%(19)
193901.


\bibitem%
%[Carmacho and Bordons 1999]
{mpc} 
%E.\ F.\ Carmacho and C.\ Bordons
Carmacho E.\ F.\ and Bordons C.\ 
(1999).
{\sl Model Predictive Control},
Springer, London.

\bibitem%
%[Claussen et al.\  1998]
{Cla98}
% J.C.\ Claussen, T.\ Mausbach, A.\ Piel, H.G.\ Schuster 
Claussen J.\ C., Mausbach T., Piel A.\ and Schuster H.\ G.\
(1998).
{\sl Phys. Rev. E} {\bf 58}, 7256.

\bibitem%
%[Claussen 1998]
{claussenthesis}
%J. C. Claussen
Claussen J.\ C.\
(1998).
{\sl Stabilisierung zeitverz\"ogert gemessener Systeme}
(in german),
% Chri\-stian-\-Al\-brechts-\-Uni\-ver\-si\-t\"at
PhD thesis,
 Kiel.


\bibitem%
%[Claussen 2002]
{claussenfloquet}
%J.C.Claussen 
Claussen J.\ C.\
(2002).
{\sl Floquet Stability Analysis of Ott-Grebogi-Yorke and Difference
Control,}
{\sl arXiv.org
e-print
} nlin.CD/0204060


\bibitem{claussenfloquetenoc}
Jens Christian Claussen, 
Influence of the impulse length in Poincare-based chaos control, Proc. ENOC 2005, Eindhoven (2005), p. 1100-1107.


%\bibitem%
%%%%%%[Claussen 2003]
%{Cla03} 
%%%%%%J.C.Claussen
%Claussen J.\ C.\
%(2003).
%Proc. 2003 Int. Conf. on Physics and Control, St.
%Petersburg, Russia, edited by 
%%%%%%%A. L. Fradkov and A. N. Churilov 
%Fradkov Al.\ L.\ and Churilov A.\ N.\ 
%(IEEE, St. Petersburg, 2003),
%Vol. 4, pp. 1296-1302.

\bibitem%
%[Claussen 2004]
{Cla04} 
%J.C.Claussen
Claussen J.\ C.\
(2004).
{\sl Phys. Rev. E} {\bf 70}, 046205.

\bibitem%
%[Claussen and Schuster 2004]
{ClaSchu04} 
%J.C.Claussen and H.G. Schuster
Claussen J.\ C.\ and Schuster H.\ G.\
(2004).
 {\sl Phys. Rev. E} {\bf 70}, 056225.

\bibitem%
%[Edwards and Spurgeon 1998]
{slidingmode}
%Christopher Edwards, Sarah K. Spurgeon
Edwards C.\ and Spurgeon S.\ K.\
(1998).
{\sl
Sliding mode control:
theory and applications},
Taylor \& Francis, London.

\bibitem%
%[Franceschini et al.\ 1999]
{franc99} 
%G. Franceschini, S. Bose, and E. Sch\"oll
Franceschini G., Bose S.\ and Sch\"oll E.
(1999).
{\sl Phys. Rev. E} {\bf 60}, 5426.


\bibitem%
%[H\"agglund 1996]
{hagglund96}
%T. H\"agglund
H\"agglund T.\
(1996).
 An industrial dead-time compensating PI controller,
{\sl Control Engineering Practice} {\bf 4}, 749-756.





\bibitem%
%[Hale 1993]
{hale} 
% J.\ K.\ Hale and S.\ M.\ Verduyn Lunel,
Hale J.\ K., Verduyn Lunel S.\ M.\ 
(1993).
{\sl Introduction to Functional Differential Equations},
Springer, New York.



\bibitem%
%[H\"ovel and  Socolar 2003]
{hovel}
%Philipp H\"ovel and Joshua E. S. Socolar
H\"ovel P.\ and  Socolar E.\ S.\
(2003).
% Stability domains for time-delay feedback control with latency
{\sl Phys. Rev. E} {\bf 68}, 036206.


\bibitem%
%[H\"ubler 1989]
{hubler} 
%A. W. H\"ubler
H\"ubler A.\ W.\ 
(1989).
{\sl
Helv. Phys. Acta} {\bf 62}, 343-346.



\bibitem%
%[Jury and Blanchard 1961]
{jury}
%E. I. Jury and J. Blanchard
Jury E.\ I.\ and Blanchard J.\
(1961).
% A Stability Test for Linear Discrete Systems in Table Forms
Proc. IRE {\bf 49}, 1947-1948;
%E. I. Jury and J. Blanchard
Jury E.\ I.\ and Blanchard J.\
(1961).
% A Stability Test for Linear Discrete Systems Using a Simple Division
Proc. IRE {\bf 49}, 1948-1949;
%
Katsuhiko Ogata, {\it Discrete-Time Control Systems},
Prentice Hall, Hertfordshire 1995;
%
J. L. Martins de Carvalho, 
{\it Dynamical Systems and Automatic Control},
Prentice Hall, Hertfordshire 1993.


\bibitem%
%[Just et al. 1997]
{just97}
%W. Just,  T. Bernard, M. Ostheimer, E. Reibold, and H. Benner
Just W., Bernard T., Ostheimer M., Reibold E. and  Benner H.\
(1997).
{\sl Phys. Rev. Lett.} {\bf 78}, 203.

\bibitem%
%[Just, Renckwerth et al. 1999]
{just99pre} 
%W. Just, D. Reckwerth, E. Reibold, and H. Benner
Just W.,  Reckwerth D, Reibold E.\ and Benner H.\
(1999).
{\sl Phys. Rev. E} {\bf 59}, 2826.

\bibitem%
%[Just, Reibold et al. 1999]
{just99pla} 
%W. Just, E. Reibold, K. Kacperski, P. Fronczak, and J. Ho{\l}yst
Just W., Reibold E.,  Kacperski K.,  Fronczak P. and Ho{\l}yst L.\
(1999).
{\sl Phys. Lett. A} {\bf 254}, 158.
% J. Holyst

\bibitem%
%[Just et al.\ 2003]
{just03pre} 
% W. Just, S. Popovich, A. Amann, N. Baba, and E. Sch\"oll
Just W., Popovich S.,   Amann A., Baba N. and Sch\"oll E.\
(2003).
{\sl Phys. Rev. E} {\bf 67}, 026222.

\bibitem%
%[Mausbach et al.\ 1997]
{mausbach95}
%Th. Mausbach, Th. Klinger, A. Piel, A. Atipo, Th. Pierre, and G. Bonhomme
Mausbach T., Klinger T., Piel A., Atipo A.,  Pierre T. and Bonhomme G.\
(1997).
%Continuous control of ionization wave chaos by spatially derived feedback signals
{\sl Phys. Lett. A} {\bf 228}, 
373.
%373--377.
%%% \\ physics/9701020


\bibitem%
%[Mausbach et al.\ 1999]
{mausbachpp99}
%T. Mausbach, T. Klinger, and A. Piel
Mausbach T.,  Klinger T.\ and Piel A.\
(1999).
Chaos and chaos control in a strongly driven thermionic plasma diode,
{\sl Physics of Plasmas} {\bf 6} 
(10),
3817-3823.



\bibitem%
%[Ott, Grebogi, Yorke 1990]
{ogy90}
Ott E., Grebogi C. and Yorke J.\ A.\
%E. Ott, C. Grebogi and J. A. Yorke
(1990).
{\sl Phys. Rev. Lett.} {\bf 64}, 1196--1199.

\bibitem%
%[Palmor 1980]
{palmor80}
Palmor Z.\
(1980).
Stability properties of Smith dead-time compensator controllers,
Int. J. Control {\bf 32} (4), 937-949.


\bibitem%
%[Petrov and Showalter 1996]
{showalter96}
%Valery Petrov and Kenneth Showalter
%V.\ Petrov and K.\ Showalter
 Petrov V.\ and  Showalter K.\
(1996).
% {\it Nonlinear Control of Dynamical Systems from a Time Series},
{\sl Phys. Rev. Lett.} {\bf 76}, 3312-3315.
% (1996)
%and references therein.


\bibitem%
%[Pyragas 1992]
{pyragas92} 
%K. Pyragas
Pyragas K.\
(1992).
% {\it Continuous control of chaos by self-controlling feedback},
{\sl Phys. Lett. A} {\bf 170}, 421--428.
% (1992).

\bibitem%
{schusterbuch} 
Schuster 
H.\ G.\
and Just W.,
Deterministic chaos: an introduction.
4.\ ed., 
Wiley-VCH, Weinheim  2005.



\bibitem%
%[Sieber and Krauskopf 2004]
{sieber04}
% J. Sieber and B. Krauskopf
Sieber J.\ and Krauskopf B.\
(2005).
% Extending the permissible control loop latency for the controlled inverted pendulum,
%J. Sieber, B. Krauskopf:  Extending the permissible control loop latency for the controlled inverted pendulum . 
Dynamical Systems
%: An international Journal 
{\bf 20},
% (2), pp. 
189-199;
%, 2005.
%
%
%
%J. Sieber and B. Krauskopf
Sieber J.\ and Krauskopf B.\
(2004).
% Bifurcation analysis of an inverted pendulum with delayed feedback control near a triple-zero eigenvalue singularity, 
{\sl Nonlinearity} {\bf 17}(1), 85-103.

\bibitem%
%[Socolar et al. 1994]
{socolar94}
%J. E. S. Socolar, D. W. Sukow, and  D. J. Gauthier
Socolar J.\ E.\ S., Sukow D.\ W.\ and  Gauthier D.\ J.\
(1994).
{\sl Phys. Rev. E} {\bf 50} (6), 3245-3248.


\bibitem%
%[Socolar and Gauthier 1998]
{socolar98}
%J. E. S. Socolar, D. J. Gauthier
Socolar J.\ E.\ S.\ and  Gauthier D.\ J.\ 
(1998).
{\sl Phys. Rev. E} {\bf 57} (6), 6589-6595.

\bibitem%
%[Smith 1953]
{smith57} 
%O.\ J.\ M.\ Smith
Smith O.\ J.\ M.\ 
(1953).
Closed control of loops with dead time,
{\sl Chemical Engineering Progress} {\bf 53},
May,
217-219. 







%\bibitem{supplement}Mathematica code for the Jury calculations
%and figures are available at the end of file\\
%{\tt http://www.theo-physik.uni-kiel.de/thesis/claussen98b.tex}.
  
%\end{references}
\end{thebibliography}
\end{document}